\documentclass[aps,twocolumn,amsmath,amssymb,preprintnumbers,floatfix,prb,superscriptaddress,longbibliography]{revtex4-2}

\usepackage{comment}
\usepackage[version=4]{mhchem}
\usepackage[utf8]{inputenc}
\usepackage{newtxtext}
\usepackage{microtype}
\usepackage{textcomp}
\usepackage{dsfont}
\usepackage{eucal}
\usepackage{siunitx}
\usepackage{soul}
\usepackage{tikz}

\usepackage{enumerate}
\usepackage{amsfonts}
\usepackage{color}
\usepackage{soul}

\usepackage{todonotes}
\presetkeys%
    {todonotes}%
    {inline}{}

\usepackage{graphicx}

\usepackage[colorlinks,allcolors=blue]{hyperref}
\usepackage[capitalize]{cleveref} 
\usepackage{cleveref}

\newcommand{\anticomm}[2]{\left\{{#1}\,,\,{#2}\right\}}                    

\newcommand{\vk}{\boldsymbol{k}}

\newcommand{\vm}{\boldsymbol{m}}

\renewcommand{\v}[1]{\boldsymbol{#1}}                           

\newcommand{\vecr}{\boldsymbol{r}}      

\definecolor{DarkBlue}{rgb}{0,0,0.80}
\definecolor{DarkRed}{rgb}{0.80,0,0}
\definecolor{Purple}{rgb}{0.55,0,0.55}
\definecolor{Purple}{rgb}{0.8,0,0.8}

\newcommand{\ie}{i.e.\ }
\newcommand{\eg}{e.g.\ }

\let\epsilon\varepsilon

\begin{document}

\title{Visualization of spin-splitter effect in altermagnets via non-equilibrium Green functions on a lattice}
\author{Karl Bergson Hallberg}
\affiliation{Center for Quantum Spintronics, Department of Physics, Norwegian \\ University of Science and Technology, NO-7491 Trondheim, Norway}
\author{Erik Wegner Hodt}
\affiliation{Center for Quantum Spintronics, Department of Physics, Norwegian \\ University of Science and Technology, NO-7491 Trondheim, Norway}
\author{Jacob Linder}
\affiliation{Center for Quantum Spintronics, Department of Physics, Norwegian \\ University of Science and Technology, NO-7491 Trondheim, Norway}\date{\today}
\begin{abstract}
When a charge current is injected into an altermagnet along a suitable crystallographic direction, a transverse spin current can be generated. This so-called spin-splitter effect does not rely on spin-orbit coupling, and is thus distinct from the spin Hall effect. The spin-splitter effect was predicted by \textit{ab initio} calculations and has been experimentally confirmed. To utilize the spin-splitter effect for practical purposes in spintronic devices, it is important to understand (i) how the system parameters affect the transverse spin current, such as filling fraction, altermagnetic strength, interface parameters, spin-orbit interactions, and impurities and (ii) determine the properties of any associated spin accumulation, which is the measurable quantity. Here, we determine the answer to these questions and provide a real-space visualization of the spin flow and spin accumulation due to the spin-splitter effect. We utilize the non-equilibrium Keldysh Green function method on a 2D square lattice to this end. We find that the presence of edges induces oscillations in the spin accumulation and strongly modify the signal for small samples. At half-filling, the spin accumulation acquires an anomalous pattern and the spin-splitter effect vanishes. We prove analytically that this follows from a combined particle-hole and spin-reversal symmetry of the model used for the altermagnetic state. Increasing the altermagnetic strength leads to a larger spin accumulation, as expected. However, when adding Rashba spin-orbit interaction, providing an additional spin Hall signal, we find that the spin accumulation is not simply the sum of the spin Hall and spin-splitter contribution. Finally, we show that the spin-splitter effect is robust towards moderate impurity scattering with a potential of the same order as the hopping parameter, which facilitates its observation in real materials.
\end{abstract}

\maketitle
\section{Introduction}

Recently, a new class of magnetic materials dubbed altermagnets have attracted significant attention \cite{Noda-Nakamura-MomentumdependentBandSpin-2016, Smejkal-Sinova:2020, Hayami-Kusunose-MomentumDependentSpinSplitting-2019, Ahn-Kunes:2019, Yuan-Zunger:2020, Yuan-Zunger-PredictionLowZCollinear-2021, Smejkal-Jungwirth:2022, Smejkal-Jungwirth-ConventionalFerromagnetismAntiferromagnetism-2022, Mazin:2022}. Despite being magnetically compensated similar to antiferromagnets, they feature time-reversal symmetry breaking and spin-split band structure similar to ferromagnets \cite{vsmejkal2022beyond}. Many material candidates such as $\mathrm{RuO_2}$, $\mathrm{CrSb}$, $\mathrm{KRu_4O_8}$, $\mathrm{MnTe}$ and $\mathrm{La_2CuO_4}$ have been suggested \cite{am_emerging} and experimental verification of spin-split electron bands and other theoretically predicted properties followed quickly \cite{krempasky_arxiv_23, bai2024altermagnetism}. The spin-polarized itinerant electrons renders the spin degree of freedom accessible for use in spintronics \cite{zarzuela_arxiv_24}.

Transport properties such as efficient charge-to-spin conversion are of high interest for next-generation spintronic devices. Earlier attempts has involved using relativistic spin-orbit based spin Hall effects to generate such transverse spin currents \cite{sinova_rmp_15, nikolic_prb_06}. Unlike conventional spin-orbit based transport mechanisms that rely on relativistic spin-orbit coupling, a spin-splitter effect in altermagnets \cite{gonzalez2021efficient, bai_prl_22, karube_prl_22} 
 arises from the spin-dependent anisotropy of the crystallographic environment. 
This generates transverse spin currents without relying on relativistic effects which may be weak compared to other interactions in the system.
Whilst altermagnets feature vanishing net magnetization and no stray magnetic fields, interface-induced effects have been shown to generate a magnetization close to the edges and vacuum interfaces \cite{hodt2024interface}. It is presently not known how this interface-induced magnetization affects the spin-splitter effect.

\textit{Ab initio} calculations for $\mathrm{RuO_2}$ \cite{gonzalez2021efficient} have shown that large spin-currents can be induced in materials hosting collinear antiferromagnetism without relativistic spin-orbit coupling as a consequence of the spin-split energy bands. These findings (which are currently under debate for $\mathrm{RuO_2}$) have created further interest in understanding both the exact theoretical mechanisms and the parameters that influence the spin-splitter effect. Recent results include investigating the charge currents and angle dependence of the transverse spin-currents \cite{wei_prb_24}, transverse spin-currents and spin accumulation in the diffusive limit \cite{kokkeler_arxiv_24} and mesoscale transport of spin and charge \cite{zarzuela_arxiv_24}.
Interestingly, spatial separation of spin carriers in the form of spin-polarized Cooper pairs is also possible using altermagnets \cite{giil_prbl_24}.

While studies have analytically and phenomenologically addressed aspects of the spin-splitter effect, a comprehensive fully quantum mechanical approach to the computation of spin accumulation as a consequence of the spin-splitter effect stemming from the band structure remains unexplored. Spin accumulation plays a crucial role in determining the feasibility of utilizing altermagnets in practical applications, as it is a measurable observable and also directly influences other measurable quantities such as magnetoresistance \cite{wei_prb_24}. Addressing this gap, our study provides a detailed quantum mechanical treatment of spin accumulation and its dependence on experimentally accessible quantities such as type of altermagnet, bias voltage, chemical potential of the leads, altermagnet strength and impurities. We provide numerically calculated values and spatially resolved plots showing the dependence of spin accumulation on these parameters.

In this paper, we utilize the non-equilibrium Keldysh Green function formalism to provide a real-space visualization of the spin flow and spin accumulation  associated with the spin-splitter effect in altermagnets. Our work is based on a tight-binding model on a two-dimensional square lattice connected to four semi-infinite leads, allowing for an in-depth investigation of how system parameters influence spin accumulation under an applied voltage bias. 

We structure this work as follows. In Sec. \ref{sec:theory} the model and theory is introduced, and we detail how the tight-binding model is set up and how to use the Keldysh Green function formalism to calculate observables. Technical details related to the calculation of the self-energy terms are presented in Appendix \ref{app:self-energies}. In Sec. \ref{sec:results} we present and discuss our results and the effect of changing the various parameters of the model, as well as the effect of including impurity scattering. In Sec. \ref{sec:conclusion} we summarize our work and present some future challenges. Throughout this article, we use units where $\hbar = e = k_B = 1$.

\begin{figure}[t!]
    \centering
\includegraphics[width=1.0\linewidth]{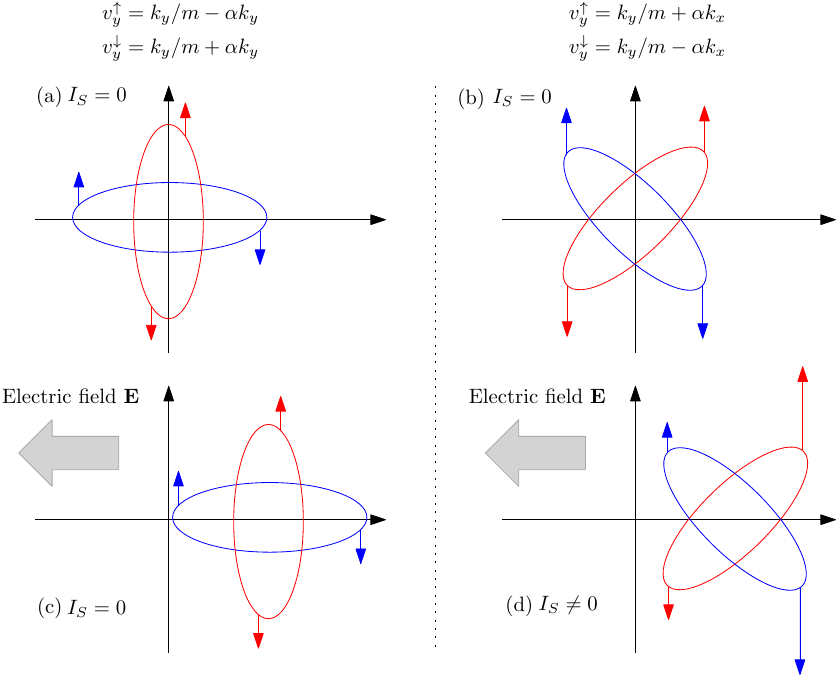}
    \caption{Illustration of the microscopic mechanism behind the spin-splitter effect. The red and blue bands are the spin-$\uparrow$ and spin-$\downarrow$ bands in the altermagnet. In (a) and (c), bands and group velocities $v_y = \partial H/\partial k_y$ for a Hamiltonian $H = k^2/2m + \alpha\sigma_z(k_x^2-k_y^2)$ are shown. In equilibrium (a), no spin current exists. When an electric field is applied along the $x$-axis, shown in (c), the Fermi-surface is displaced. Yet, no transverse spin current flows because the group velocities of the particles in the $y$-direction remain unchanged and compensate each other on each individual Fermi surface. In (b) and (d), bands and group velocities for a Hamiltonian $H=k^2/2m + \alpha\sigma_zk_xk_y$ are shown. In equilibrium (b), no spin current exists. When an electric field is applied along the $x$-axis, shown in (d), the Fermi-surface is displaced. Now, a transverse spin current is generated. This is because the group velocities in the $y$-direction no longer cancel each other on each individual Fermi surface, and because there for each point on the spin-$\uparrow$ Fermi surface exists a point with opposite group velocity on the spin-$\downarrow$ Fermi surface.}
    \label{fig:mechanism}
\end{figure}

\section{Theory}\label{sec:theory}
\begin{figure}
    \centering
    \includegraphics[width=1.0\linewidth]{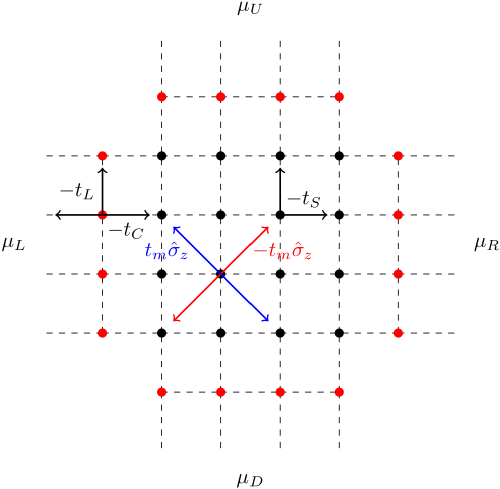}
    \caption{Schematic figure showing a central $d_{xy}$-altermagnet lattice (pictured with black dots) connected to four leads (pictured with red dots). The leads are connected at infinity to reservoirs in equilibrium with a chemical potential of $\mu_\ell$.}
    \label{fig:schematiclattice}
\end{figure}

To model the 2D $N_x \times N_y$ lattice connected to four semi-infinite leads we employ a tight-binding model with a Hamiltonian given by
\begin{equation}\label{hamiltonian}
    \hat H = \sum_{\vm \sigma} \epsilon_{\vm} \hat{c}^\dagger_{\vm \sigma} \hat{c}_{\vm \sigma} + \sum_{\vm \vm'\sigma \sigma'}\hat{c}^\dagger_{\vm \sigma}t_{\vm \vm'}^{\sigma \sigma'} \hat{c}_{\vm' \sigma'},
\end{equation}
with $\hat{c}^\dagger_{\vm \sigma}$ ($\hat{c}_{\vm \sigma}$) creating (annihilating) an electron with spin $\sigma$ on the site at $\vm$. The on-site energy $\epsilon_{\vm} $ may be used to simulate a static local potential or disorder in the system if chosen as a random variable. The altermagnet central region is modeled by setting the hopping parameters $t_{\vm \vm'}$ for $\vm, \vm'$ both belonging to the central sample region as $2 \times 2$ matrices acting in spin space,
\begin{equation}\label{xy am hopping}
    t_{\vm \vm'} = \begin{cases}
        -t_S \v I, \quad (\vm = \vm' \pm \v e_x \; \mathrm{or} \; \vm = \vm' \pm \v e_y), \\
        -t_m \hat \sigma_z, \quad (\vm = \vm' \pm (\v e_x + \v e_y)), \\
        t_m\hat \sigma_z, \quad (\vm = \vm' \pm (\v e_x - \v e_y)),
    \end{cases}
\end{equation}
for a $d_{xy}$-altermagnet and
\begin{equation}\label{parallel am hopping}
    t_{\vm \vm'} = \begin{cases} -t_S \v I - t_m \hat \sigma_z, \quad (\vm = \vm' \pm \v e_y), \\
    -t_S \v I + t_m \hat \sigma_z, \quad (\vm = \vm' \pm \v e_x),
    \end{cases}
\end{equation}
for a $d_{x^2-y^2}$ altermagnet (see Fig. \ref{fig:schematiclattice}).

Here $\v I$ is the spin space identity matrix. The leads are modeled as clean metals with $\epsilon_{\vm} = 0$ and 
\begin{equation}\label{lead hopping}
    t_{\vm \vm'}^L = -t_L \v I,
\end{equation}
for $\vm, \vm'$ nearest neighbors both belonging to lattice points on the leads. For the interface between the leads and sample region we only consider nearest neighbor hopping with
\begin{equation}
    t_{\vm \vm'}^C = - t_C \v I,
\end{equation}
for \eg $\vm$ belonging to the outermost row/column of the lead adjacent to $\vm'$ corresponding to the outermost row/column of the central sample region.

At infinity, the four leads are connected to reservoirs with constant chemical potentials $\mu_L, \mu_R, \mu_U$ and $\mu_D$ respectively. The left and right reservoirs are biased relative to each other by a non-zero $eV = \mu_L - \mu_R$, and electrons distributed according to equilibrium Fermi-Dirac with chemical potential of the reservoirs are injected into the system. The reservoirs are assumed to be large enough that they, and the leads, are kept in equilibrium and the system is in a steady state.

The full Hamiltonian of the system may be written $\hat H = \hat H_S + \hat H_\mathrm{leads}+ \hat H_C$ for the sample, leads and the connection between the sample and leads respectively. In the occupation number basis for the sites $\vm$ and spin $\sigma$, the sample region Hamiltonian $\hat H_S$ is a finite $2N_xN_y \times 2N_xN_y$ matrix, while $\hat H_\mathrm{leads} = \hat H_L + \hat H_R + \hat H_U + \hat H_D$ and $\hat H_C$ are infinite. The two-dimensional central sample region can be indexed by a single number via the correspondence
\begin{equation}
    (m_x, m_y) \longrightarrow i = m_y + N_ym_x,
\end{equation}
which runs from $i = 0$ to $N_xN_y - 1$. 

To calculate non-equilibrium statistical averages we make use of the Keldysh Green-function technique. This entails assuming that in the infinite past the system is in thermal equilibrium with density matrix given by
\begin{equation}
    \hat \rho(-\infty) = \frac{1}{Z} \big ( \hat \rho_S^{(0)} \otimes e^{-\beta (\hat H_L - \mu_L N_L)} \otimes \dots \otimes e^{-\beta ( \hat H_D - \mu_D N_D )} \big ).
\end{equation}
In other words, in the infinite past the leads and sample are unconnected, the leads each in thermal equilibrium with chemical potential $\mu_\ell$, $\ell \in\{ L,R,U,D\}$, whereas $\hat \rho_S^{(0)}$ is the density matrix of the central region in the infinite past. The chemical potential of the central region in the infinite past has no bearing on the results in the steady state Keldysh formalism and is not included in the Hamiltonian. This is because of an assumption that the interaction (electron hopping) between the leads and the sample has been turned on adiabatically and over a sufficiently long time that the initial condition specific for the uncorrelated initial state on the sample has become irrelevant \cite{keldysh1965diagram, maciejko2007nonequilibrium}. This is manifested mathematically in that one derives the so-called Keldysh equation for the lesser Green function in the central region without any knowledge of the specific initial state of the central region except that it was in equilibrium \cite{spicka_ijmpb_14} in the infinite past. In effect, we are neglecting any transient phenomena related to turning on a coupling between the leads and central region.
As the central region is driven out of equilibrium and reaches its steady-state, the occupation of electron states is not described by a Fermi-Dirac distribution, but rather determined by the thermalized leads.

Unlike equilibrium statistical mechanics, we do not assume that the final state in the infinite future differs only by a phase from the ground state in the infinite past. Instead, there is a forward and backward contour which enables calculating non-equilibrium statistical averages with respect to the equilibrium density matrix \cite{arseev2015nonequilibrium}. 

Since we are considering steady state transport, the two-point correlation functions 
\begin{equation}
    \langle \hat c^\dagger_{\vm' \sigma'}(t') \hat c_{\vm \sigma}(t) \rangle = -iG^<_{\vm \vm', \sigma \sigma'}(t, t'),
\end{equation}
depend only on the time difference $\tau \equiv t - t'$, which facilitates a Fourier transform to energy
\begin{equation}
    G^<_{\vm \vm', \sigma \sigma'}(\tau) = \frac{1}{2\pi} \int_{-\infty}^\infty dE G^<_{\vm \vm', \sigma \sigma'}(E)e^{iE\tau}.
\end{equation}
Equal time two-point functions are then given by
\begin{equation}
    \langle \hat c^\dagger_{\vm \sigma} \hat c_{\vm' \sigma'} \rangle =\frac{1}{2 \pi i} \int_{-\infty}^\infty dE \, G^<_{\vm' \vm, \sigma' \sigma}(E).
\end{equation}
The local charge density is given as the statistical average of the electron number operator for site $\vm$
\begin{equation}
\begin{split}
    \langle \hat N_m \rangle &= \sum_{\sigma} \langle \hat c^\dagger_{\vm \sigma} \hat c_{\vm \sigma} \rangle = \sum_{\sigma} \frac{1}{2\pi i}\int_{-\infty}^\infty dE G^<_{\vm \vm, \sigma \sigma}(E) \\
    &= \frac{1}{2\pi i}\int_{-\infty}^\infty dE \; \mathrm{tr_s}G^<_{\vm \vm} (E), \label{eqn:poisson}
\end{split}
\end{equation}
where $\mathrm{tr_s}$ is a trace over the $2 \times 2$ sub-matrix $G^<_{\vm \vm}$ containing the spin degrees of freedom of site $\vm$. 

The spin-resolved currents may be attained by using conservation of charge and the Heisenberg equation of motion for the change in electron number at site $\vm$,
\begin{equation}
     \frac{d \hat{N}_{\vm}}{dt} = -i [ \hat{N}_{\vm}, \hat{H} ].
\end{equation}
Following \cite{nikolic_prb_06}, we introduce the bond charge-current operator $\hat J_{\vm \vm'}$ which represents the particle current from site $\vm$ to $\vm'$. For normal nearest neighbor hopping and $d_{x^2-y^2}$-altermagnets the sites $\vm$ and $\vm'$ are nearest neighbors, but include next-nearest neighbors for the case of an $d_{xy}$-altermagnet. Local charge conservation on the lattice then takes the form
\begin{equation}
    \frac{d \hat N_{\vm}}{dt} + \sum_{\vm'}(\hat J_{\vm \vm'} - \hat J_{\vm' \vm}) = 0.
\end{equation}
The bond charge-current operator may be spin-resolved by using Eq. (\ref{xy am hopping}) or (\ref{parallel am hopping}) as
\begin{equation}
    \hat J_{\vm \vm'} = \sum_{\sigma \sigma'} \hat J_{\vm \vm'}^{\sigma \sigma'} = -i \sum_{\sigma \sigma'} [\hat c^\dagger_{\vm' \sigma'} t_{\vm' \vm}^{\sigma' \sigma} \hat c_{\vm \sigma} - \mathrm{h.c.}].
\end{equation}
The statistical average of the spin-resolved bond charge-current operator is, in steady state after transients have died away, given by
\begin{align}
    \langle \hat J_{\vm \vm'}^{\sigma \sigma'} \rangle = -\frac{1}{2 \pi} \int_{-\infty}^\infty dE \, [&t_{\vm' \vm}^{\sigma' \sigma}G^<_{\vm \vm', \sigma \sigma'}(E) \notag\\
    - &t_{\vm \vm'}^{\sigma \sigma'}G^<_{\vm' \vm, \sigma' \sigma}(E)].
\end{align}
Inserting $t_{\vm \vm'}^{\sigma \sigma'}$ from Eq. (\ref{xy am hopping}) yields, for a $d_{xy}$-altermagnet
\begin{equation}
\begin{split}
    \langle \hat J_{\vm \vm'} \rangle = \frac{t_S}{2 \pi}\int_{-\infty}^\infty dE \; \mathrm{tr_s}[G^<_{\vm \vm'}(E) - G^<_{\vm' \vm}(E)] \\
    \times (\delta_{\vm', \vm \pm \v e_x} + \delta_{\vm', \vm \pm \v e_y}) \\
    + \frac{t_m}{2\pi}\int_{-\infty}^\infty dE \; \mathrm{tr_s}(\hat \sigma_z[G^<_{\vm \vm'}(E) - G^<_{\vm' \vm}(E)])\\
    \times (\delta_{\vm', \vm \pm (\v e_x - \v e_y)} - \delta_{\vm', \vm \pm (\v e_x + \v e_y)}).
\end{split}
\end{equation}
Similarly for the bond spin-current, we follow \cite{nikolic_prb_06} and introduce the symmetrized product of the spin-$1/2$ operator $\hat \sigma_i/2$ and the bond charge-current operator,
\begin{equation}
    \hat J^{S_i}_{\vm \vm'} = \frac{1}{4i}\sum_{\alpha \beta}(\hat c^\dagger_{\vm' \beta} \anticomm{\hat \sigma_i}{t_{\vm' \vm }}_{\beta \alpha} \hat c_{\vm \alpha} - \mathrm{h.c.}).
\end{equation}
Without spin-orbit coupling, the anti-commutation rules for $\hat{\sigma}_z$ yields
\begin{equation}
\begin{split}
    \hat J^{S_z}_{\vm \vm'} = \frac{it_S}{2}\sum_{\alpha \beta} ( \hat c^\dagger_{\vm' \beta}(\hat \sigma_z)_{\beta \alpha} \hat c_{\vm \alpha} - \mathrm{h.c.}) \\
    \times(\delta_{\vm', \vm \pm \v e_x} + \delta_{\vm', \vm \pm \v e_y}) \\
    + \frac{it_m}{2}\sum_{\alpha \beta}(\hat c^\dagger_{\vm' \beta}\delta_{\beta \alpha}\hat c_{\vm \alpha} - \mathrm{h.c.}) \\
    \times (\delta_{\vm', \vm \pm (\v e_x + \v e_y)} -  \delta_{\vm', \vm \pm (\v e_x - \v e_y)}).
\end{split}
\end{equation}
In terms of the lesser Green function, the statistical average of this operator is given by
\begin{equation}\label{spincurrent}
\begin{split}
    \langle \hat J^{S_z}_{\vm \vm'} \rangle = \frac{t_S}{4\pi}\int_{-\infty}^\infty dE \; \mathrm{tr_s}[\hat \sigma_z(G^<_{\vm' \vm} - G^<_{\vm \vm'})] \\
    \times (\delta_{\vm', \vm \pm \v e_x} + \delta_{\vm', \vm \pm \v e_y}) \\
    + \frac{t_m}{4\pi} \int_{-\infty}^\infty dE \; \mathrm{tr_s}(G^<_{\vm' \vm} - G^<_{\vm \vm'}) \\
    \times (\delta_{\vm', \vm \pm (\v e_x + \v e_y)} - \delta_{\vm', \vm \pm (\v e_x - \v e_y)}).
\end{split}
\end{equation}
Finally, the components of the local spin operator is defined at site $\vm$ as
\begin{equation}
    \hat{S}^i_{\vm} = \frac{1}{2}\sum_{\alpha \beta} \hat c^\dagger_{\vm \alpha}(\hat \sigma_i)_{\alpha \beta}\hat c_{\vm \beta},
\end{equation}
which yields an expression for the statistical average of the local spin density via
\begin{equation}\label{spindensity}
    \langle \hat{S}^i_{\vm} \rangle = \frac{1}{4\pi i}\int_{-\infty}^\infty dE \; \mathrm{tr_s}(\hat \sigma_i G^<_{\vm \vm}(E)).
\end{equation}

To calculate the lesser Green function, we make use of the Keldysh equation (see Appendix \ref{sec:keldysh}) for mesoscopic transport
\begin{equation}\label{keldysheq}
    G^< = G^R\Sigma^< G^A,
\end{equation}
valid for steady-state transport a long time after transients have died away \cite{keldysh1965diagram}. Here $G^R$ and $G^A = (G^R)^\dagger$ are the retarded and advanced Green functions respectively. Naively calculating $G^{R,A}$ involves, even in the occupation number basis on the lattice, inverting an infinite matrix
\begin{equation}
    G^{R,A} = [E - \hat H_S - \hat H_{\mathrm{leads}} - \hat H_C \pm i \eta ]^{-1}.
\end{equation}
Instead of including the full Hamiltonian of the infinite leads, the interaction between the altermagnet central region and the semi-infinite leads are taken into account by self-energy terms. We proceed to show how this is done.

In the case of non-interacting electrons in the sample-region the self-energy terms in this region are exactly calculable. This is made possible by writing the Green function in the form
\begin{equation}
    G^R = \begin{pmatrix}
        G^R_{\mathrm{lead}} & G^R_C \\\\
        (G^R_C)^\dagger & G^R_{S}
    \end{pmatrix} = 
    \begin{pmatrix}
        E + i \eta - \hat{H}_{\mathrm{lead}} & \hat H_C \\\\
        \hat H_C^\dagger & E + i \eta - \hat H_S
    \end{pmatrix}^{-1},
\end{equation}
and using $(G^R)^{-1}G^R = I$, resulting in
\begin{align}
    [E + i\eta - \hat H_{\mathrm{lead}}]G^R_C + \hat H_C G^R_S = 0, \\
    \hat H_C^\dagger G^R_C + [E + i \eta - \hat H_S]G^R_S = I. \label{second matrix eq}
\end{align}
Solving the first equation for $G^R_C$ yields
\begin{align}
    G^R_C = -g^R_{\mathrm{lead}}\hat H_C G^R_S, \label{GRCeq}\\
    g^R_{\mathrm{lead}} \equiv [E + i \eta - \hat H_{\mathrm{lead}}]^{-1},
\end{align}
where $g^R_{\mathrm{lead}}$ is the Green function for a bare semi-infinite lead. Inserting Eq. (\ref{GRCeq}) into Eq. (\ref{second matrix eq}) then yields
\begin{equation}
\begin{split}
    (E - \hat H_S - \hat H_C^\dagger \, g^R_{\mathrm{lead}} \, \hat H_C)G^R_S = I, \\
    \implies G^R_S = (E - \hat H_S - \hat H_C^\dagger \, g^R_{\mathrm{lead}} \, \hat H_C)^{-1}.
\end{split}
\end{equation}
The last term in the expression for $G^R_S$ above thus takes the form of an effective self-energy. The term $\hat H_C^\dagger g^R_{\mathrm{lead}} \hat H_C \equiv \Sigma_{\mathrm{lead}}$ then provides a well-defined imaginary part to the Green function, and ensures that $G^R_S$ is a retarded function. Therefore, the extra $+i\eta$ in the definition of $G^R_S$ is unnecessary and can be left out.

The benefit of writing $G^R_S$ in this form is that, in the occupation number basis detailed previously, $\Sigma_{\mathrm{lead}}$ is a $2N_xN_y \times 2N_xN_y$ matrix we can calculate exactly. The problem of inverting an infinite matrix has therefore been reduced to calculating the Green function for bare leads and inverting a finite matrix. A detailed calculation of the self-energy terms are given in Appendix \ref{app:self-energies}.

With four leads, the expression for the retarded Green function and the lesser self-energy $\Sigma^<$ are given by
\begin{equation}\label{retgreens}
    G^R(E) = \bigg(E - \hat H_S - \sum_\ell \Sigma_\ell (E- eV_\ell) \bigg)^{-1}
\end{equation}
and
\begin{equation}\label{sigmalesser}
    \Sigma^<(E) = i \sum_{\ell}\Gamma_\ell(E - eV_\ell)f_\ell(E - eV_\ell),
\end{equation}
where $\Sigma_\ell$ and $\Gamma_\ell$ are defined by
\begin{align}
    \Gamma_\ell(E) &= i(\Sigma_\ell(E) - \Sigma^\dagger_\ell(E)), \label{Gamma} \\
    \Sigma_\ell(E) &= \hat H_C^\dagger (E + i \eta - \hat H_\ell)^{-1}\hat H_C.
\end{align}
The index $\ell = L, R, U, D$ refers to the leads situated to the left, right, above or below the altermagnet respectively. Here $f(E)$ is the Fermi-Dirac distribution measured from the Fermi-level of the leads in equilibrium, or in other words the equilibrium distribution of the electrons injected from the reservoirs.

To obtain the expression given above for $\Sigma^<$ out of equilibrium, it is instructive to first briefly consider the expression for the equilibrium density matrix in terms of the retarded Green function
\begin{equation}\label{equilibriumdensitymatrix}
    \hat{\rho}_{\mathrm{eq}} = -\frac{1}{\pi}\int_{-\infty}^\infty \mathfrak{Im} (G^R)f(E) dE \equiv \frac{1}{2\pi}\int_{-\infty}^\infty \hat{A}f(E)dE,
\end{equation}
where $\hat{A} \equiv i(G^R - G^A)$ is the spectral function. This expression follows from the fact that the density operator is generally given by the lesser Green function, which in turn is expressed via the retarded Green function and the Fermi-Dirac distribution in equilibrium. To see this, consider the expectation value of a one-particle operator $\hat{O}$:
\begin{align}
    \langle \hat{O}\rangle = \text{Tr}\{ \hat{\rho}\hat{O}\} = \sum_{ij}  O_{ij}\rho_{ji}
\end{align}
where the trace has been performed using a complete basis set $\{|i\rangle\}$, with $O_{ij}$ being the matrix elements of $\hat{O}$ in this basis set. At the same time, we can also express the operator in terms of second quantized field operators:
\begin{align}
    \hat{O}= \int d\vecr \psi^\dag(\vecr) \hat{O}\psi(\vecr) = \sum_{ij} O_{ij} \hat{c}_i^\dag \hat c_j
\end{align}
where the field operators are given by $\psi(\vecr) = \sum_i \Phi_i(\vecr) \hat c_i$ and $\Phi_i(\vecr) = \langle \vecr|i\rangle$. We thus arrive at
\begin{align}
    \langle \hat{O}\rangle = \sum_{ij}O_{ij} \langle \hat c_i^\dag \hat c_j\rangle
\end{align}
and can identify the density matrix as $\rho_{ij} = \langle \hat c_j^\dag \hat c_i\rangle$, the latter being precisely the lesser Green function defined earlier. 

The spectral function $\hat A$ may be related to the imaginary part of the self-energies, $\Gamma_\ell$, via
\begin{equation}
    (G^R)^{-1} - (G^A)^{-1} = \sum_\ell(- \Sigma^R_\ell + \Sigma^A_\ell) \equiv i \sum_\ell \Gamma_\ell.
\end{equation}
By multiplying on the left by $G^R$ and right by $G^A$, we obtain
\begin{equation}
    -i \hat A = G^A - G^R = i \sum_\ell G^R \Gamma_\ell G^A,
\end{equation}
which we may rewrite by defining lead-resolved spectral functions $\hat A_\ell \equiv G^R \Gamma_\ell G^A$ by
\begin{equation}
    \hat A = \sum_\ell G^R \Gamma_\ell G^A \equiv \sum_\ell \hat A_\ell.
\end{equation}
 The non-equilibrium density matrix in the central region may now be obtained by the ansatz that the spectral functions $\hat A_\ell$ of the leads are filled according to the distribution functions $f_\ell(E) \equiv f(E - \mu_\ell)$. This gives:
\begin{equation}
\begin{split}
    \hat{\rho}_{\mathrm{neq}} &= \frac{1}{2\pi}\int_{-\infty}^\infty \sum_\ell \hat A_\ell f_\ell(E)dE\\
    &= \frac{1}{2\pi i}\int_{-\infty}^\infty G^R \big( \sum_\ell i\Gamma_\ell f_\ell \big) G^A dE \\
    &= \frac{1}{2\pi i}\int_{-\infty}^\infty G^<(E)dE.
\end{split}
\end{equation}
From this, it follows that since $G^< = G^R \Sigma^< G^A$ according to the Keldysh equation, $\Sigma^<$ is given precisely by Eq. \ref{sigmalesser} \cite{nikoliclecturenotes}.

In writing Eq. (\ref{retgreens}), we are neglecting the effects of the potential landscape within the central region. This entails that the induced non-equilibrium charge distribution has no feedback effect on the current itself. This could be compensated for by including the non-equilibrium potential landscape in the retarded Green function, obtained through Eq. (\ref{eqn:poisson}) (multiplied by $e$). The reciprocal relation between induced density and current would have to be solved self-consistently for the actual ``screened" current in the central region. We argue, however, that by applying a bias small compared to the distance between the band bottom $E_\text{min}$ and the Fermi level $E_F$, we are in a linear-response regime and the omission of this correction thus reasonable \cite{linearresponse}.

\begin{figure}
    \centering
    \includegraphics[width=0.9\linewidth]{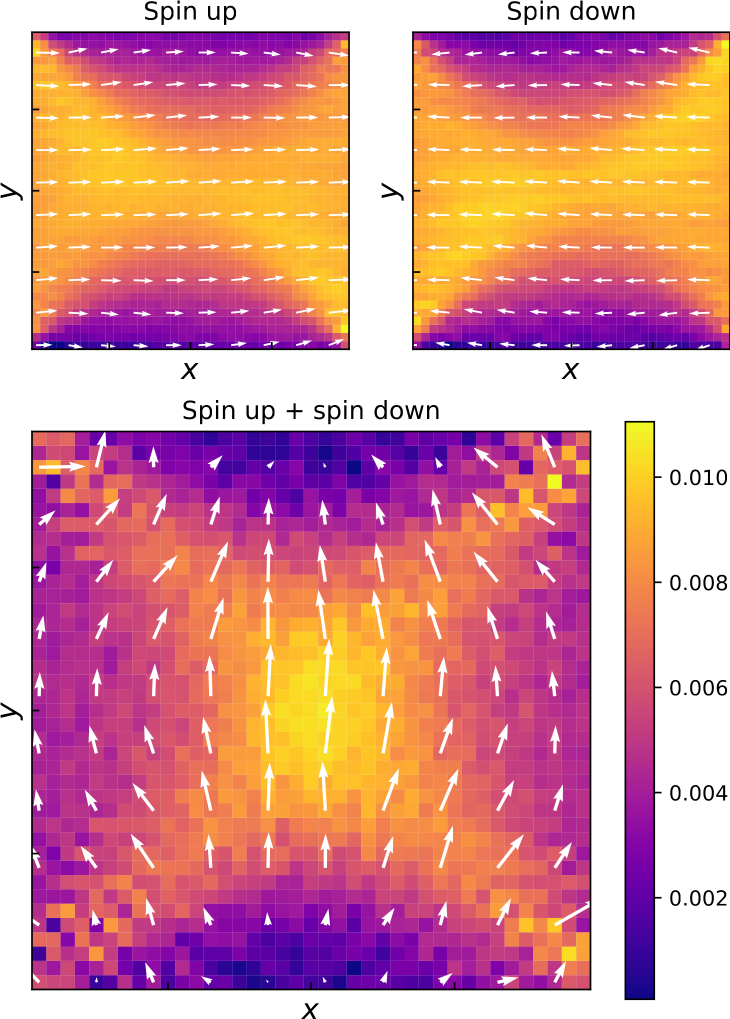}
    \caption{Direction and magnitude of the local spin currents $\langle J_{\vm \vm'}^{S_z} \rangle$ for a system with a $40\times40$ site $d_{xy}$-altermagnet central region of strength $t_m = 0.1t_S$ and four semi-infinite leads connected to reservoirs with Fermi-level $E_F = -2.0t_S$. The chemical potential of the left and right reservoirs are biased relative to each other via $\mu_{L,R} = E_F \pm eV/2$ where $eV = 0.2t_S$. The top plots show the spin-up and -down resolved currents, whilst the bottom shows the total local spin currents.}
    \label{fig:40lat0.1am-2.0fermi0.2bias-current}
\end{figure}
\begin{figure}
    \centering
    \includegraphics[width=0.9\linewidth]{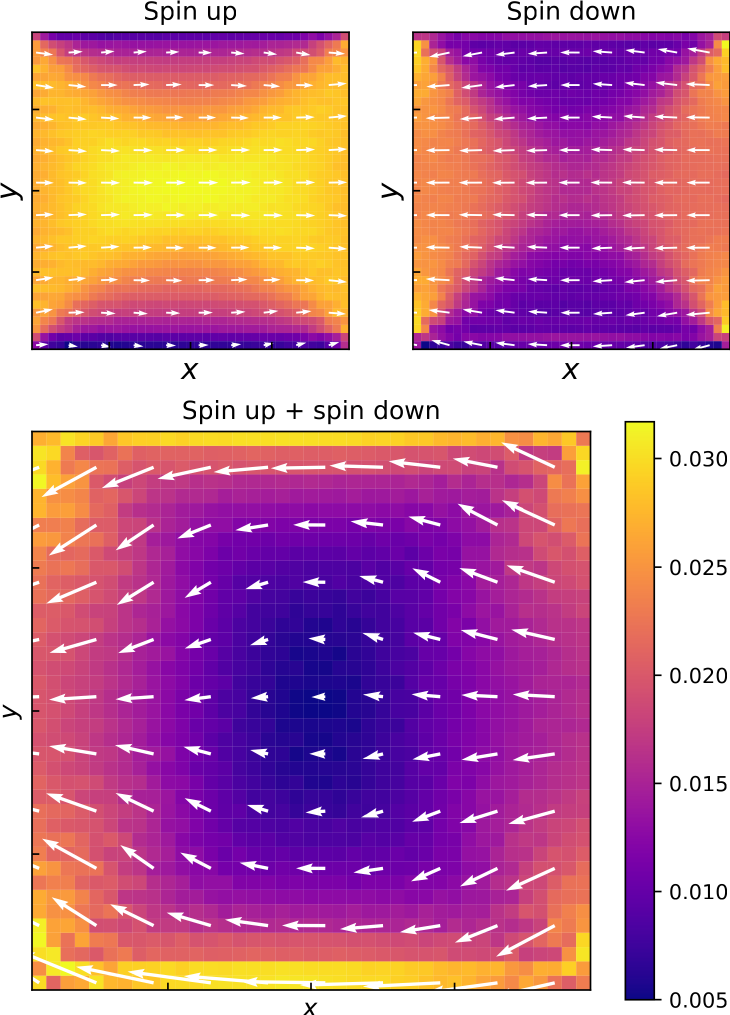}
    \caption{Direction and magnitude of the local spin currents $\langle J_{\vm \vm'}^{S_z} \rangle$ for a system with a $40 \times 40$ site $d_{x^2-y^2}$-altermagnet central region of strength $t_m = 0.1t_S$ connected to four semi-infinite leads leading to reservoirs with Fermi-level $E_F = -2.0t_S$. The chemical potential of the left and right reservoirs are biased relative to each other via $\mu_{L,R} = E_F \pm eV/2$ where $eV = 0.2t_S$. }
    \label{fig:40lat0.1am-2.0fermi0.2biasPARA-current}
\end{figure}

\section{Results and Discussion}\label{sec:results}

\subsection{Spin current}
Spin
currents are not generally conserved in systems with for
instance spin-orbit interactions or non-collinear magnetic
order \cite{shi_prl_06}, and become ambiguously defined as can be seen from the spin continuity equation. In altermagnetic systems without spin-orbit interactions, the $z$-component of the spin current is nevertheless conserved. The spin current is not measured directly. Instead, it is typically the spin accumulation (magnetization) resulting from spin flow which is experimentally probed. Nevertheless, our framework does allow for computation of the spatially resolved spin-current in an altermagnet driven out of equilibrium. We here briefly show some results for this current to give the reader an idea of how spin is transported when the system is subject to an external electric voltage.

In Fig. \ref{fig:40lat0.1am-2.0fermi0.2bias-current}, we plot the spin-resolved bond-currents (given by Eq. (\ref{spincurrent})) for a system with a $d_{xy}$-altermagnet central region under an applied bias over the left and right leads. The spin-anisotropy in the band structure leads to an anisotropy in the up- and down-components resulting in a net transversal spin current with direction depending on the sign of the applied bias. This is to be contrasted with a system with a $d_{{x^2-y^2}}$-altermagnet central region which under the same applied bias produces a longitudinal spin-polarized current in the direction of applied bias, as seen in Fig. \ref{fig:40lat0.1am-2.0fermi0.2biasPARA-current}. This is consistent with the angle-dependent results obtained in \cite{wei_prb_24}. 

In the remainder of this work, we focus on the $d_{xy}$-altermagnet where a spin-splitter effect arises. The plots for the spin accumulation are produced using a bicubic interpolation to ease viewing.

\subsection{Spin accumulation from spin-splitter effect}

\subsubsection{Interface-induced spin accumulation}\label{subsec:interfacemag}

\begin{figure}[h!]
    \centering
    \includegraphics[width=1.0\linewidth]{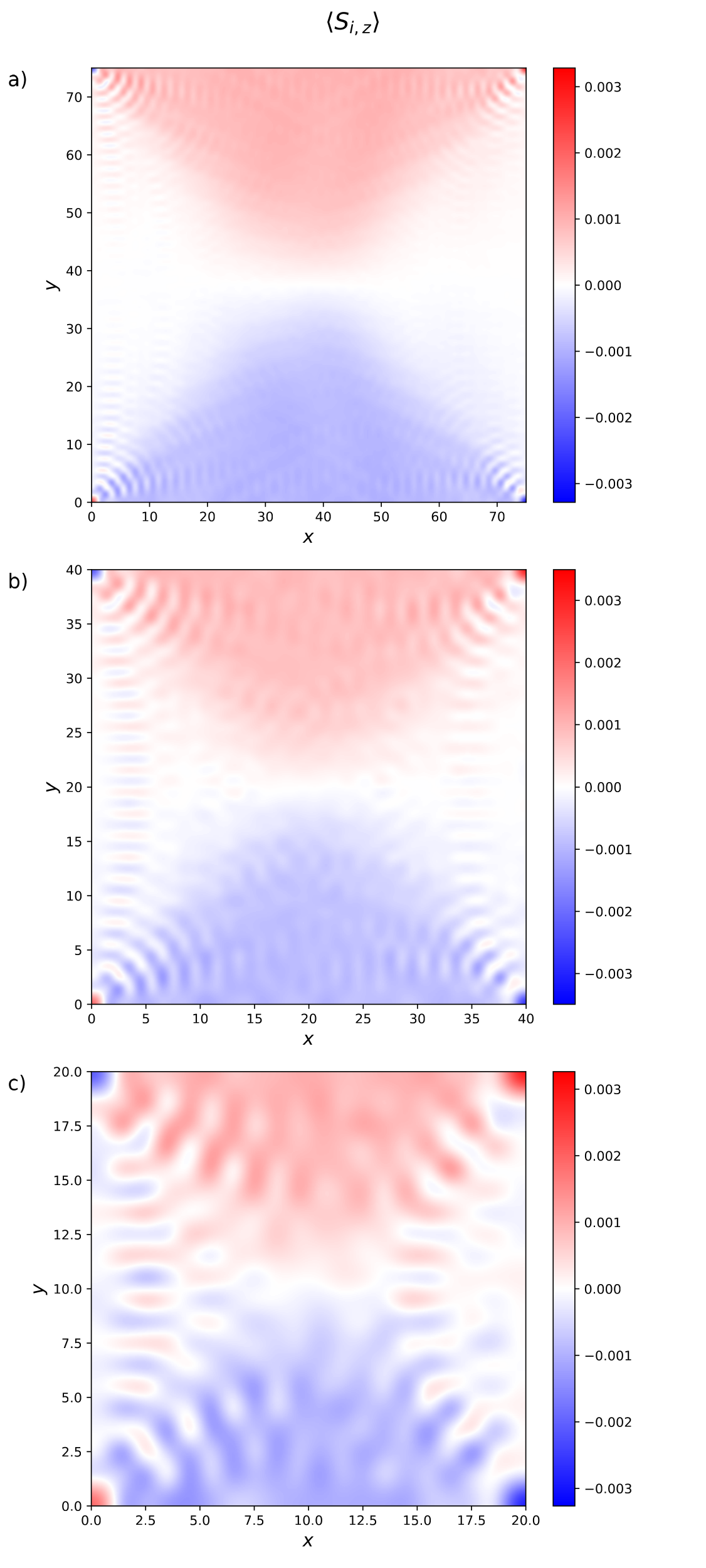}
    \caption{Local spin densities $\langle \hat S_{\vm}^z \rangle$ for a $d_{xy}$-altermagnet system with $t_m = 0.1t_S$, $E_F = -2.0t_S$ and applied bias $eV = 0.2t_S$ for a) a $75 \times 75$ site system, b) a $40 \times 40$ site system and c) a $20 \times 20$ site system. As the system size is increased past $40 \times 40$, the small lattice size effects decrease and becomes insignificant.}
    \label{fig:40lat0.1am-2.0fermi0.2bias-spindensity}
\end{figure}

In Fig. \ref{fig:40lat0.1am-2.0fermi0.2bias-spindensity} we plot the $z$-component of the spin density (given by the expression in Eq. (\ref{spindensity})) in systems with four semi-infinite leads connected on all four sides of the square lattice $d_{xy}$-altermagnet of strength $t_m = 0.1t_S$. The semi-infinite leads are connected at infinity to a reservoir of thermalized electrons following a distribution with chemical potential $\mu = -2.0t_S$.  A small bias voltage is applied over the left and right leads leading to a spin-current and spin accumulation on the interface, the strength of the bias is chosen to be 10\% of the distance from the band-bottom to the Fermi-level such as to warrant a non-selfconsistent treatment of the electric potential in the central region. In practice this means the chemical potential of the left and right reservoirs are kept at $\mu_L = E_F + eV/2$ and $\mu_R = E_F - eV/2$ as explained in Sec. \ref{sec:theory}. The upper and lower reservoirs are kept at $\mu_U = \mu_D = E_F$ here and for the remainder of the article. Introducing an additional bias between these reservoirs results  in rotating the net current relative to the altermagnet band structure and would therefore yield results known from $d_{x^2 - y^2}$-altermagnets.

\begin{figure}
    \centering
    \includegraphics[width=1.0\linewidth]{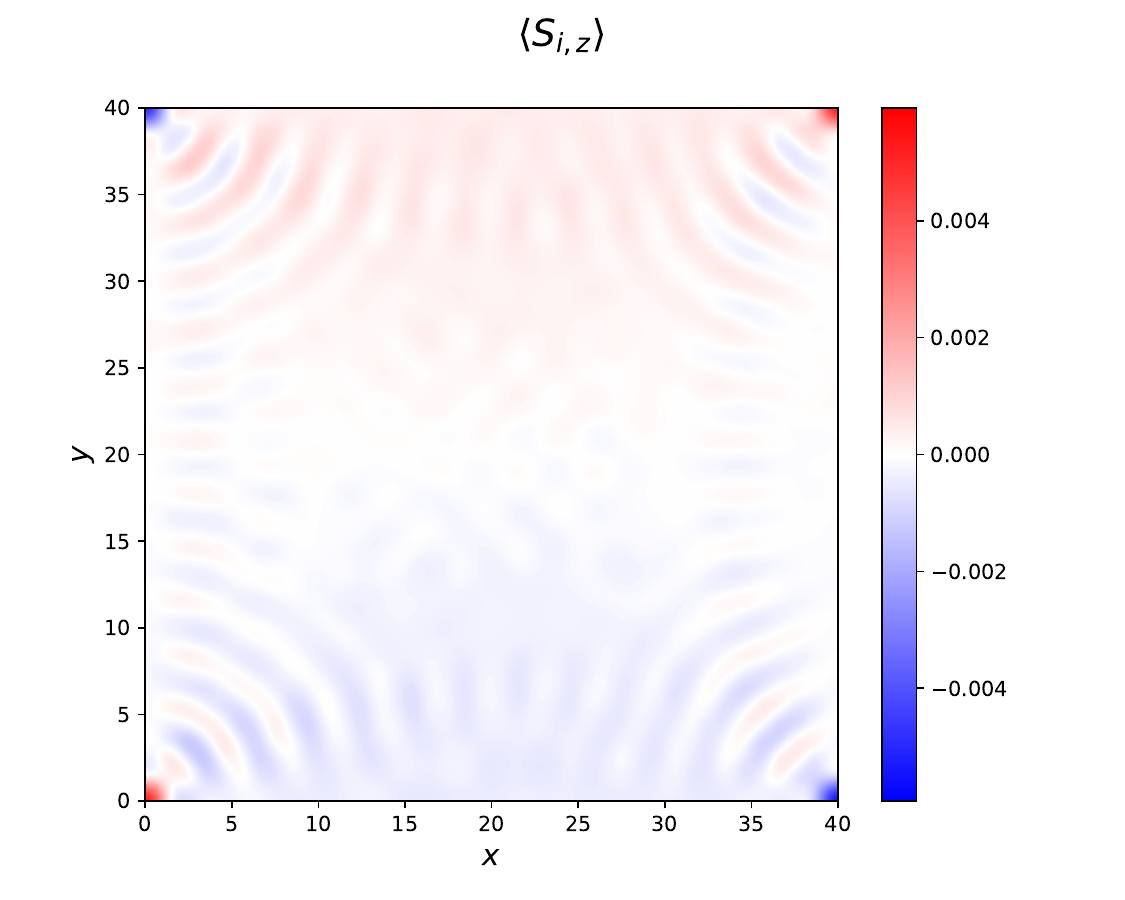}
    \caption{Local spin density $\langle \hat S_{\vm}^z \rangle$ for a $40 \times 40$ site $d_{xy}$-altermagnet system with $t_m = 0.1t_S$, $E_F = -3.0t_S$ and applied bias $eV = 0.1t_S$.}
    \label{fig:40lat0.1am-3.0fermi0.1bias-spindensity}
\end{figure}

The plot of the of the local spin density shows a clear transverse accumulation of the $z$-component at the interface between altermagnet and the upper and lower leads as expected from the statistically averaged spin current. The accumulation stays consistent when changing the values of the parameters such as lattice size or small changes of chemical potential, as seen in Fig. \ref{fig:40lat0.1am-2.0fermi0.2bias-spindensity} and Fig. \ref{fig:40lat0.1am-3.0fermi0.1bias-spindensity} for different lattice sizes and for $E_F = -3.0t_S$ respectively. As an estimate for the magnetic moment per area of the film for experimental purposes, for a $40\times 40$ system with $t_m = 0.1t_S$, $E_F = -2.0t_S$ and $eV = 0.2t_S$, a rough mean value for each site is $0.001 \hbar$. Measuring in the area with largest spin accumulation should then yield circa $0.001\hbar/a^2$, or a magnetic moment per area of $0.001g\mu_B/a^2$. Using $g\approx2$ and $a=5\;\r{A}$, the experimentally available signal should be at least $0.002 \cdot \mu_B /(2.5 \cdot 10^{-19} \mathrm{m^2}) \approx 8 \cdot 10^{15} \mu_B/m^2$.  The current-induced spin accumulation can be experimentally probed using techniques such as the magneto-optic Kerr effect or SQUID measurements.

\begin{figure}
    \centering
    \includegraphics[width=1.0\linewidth]{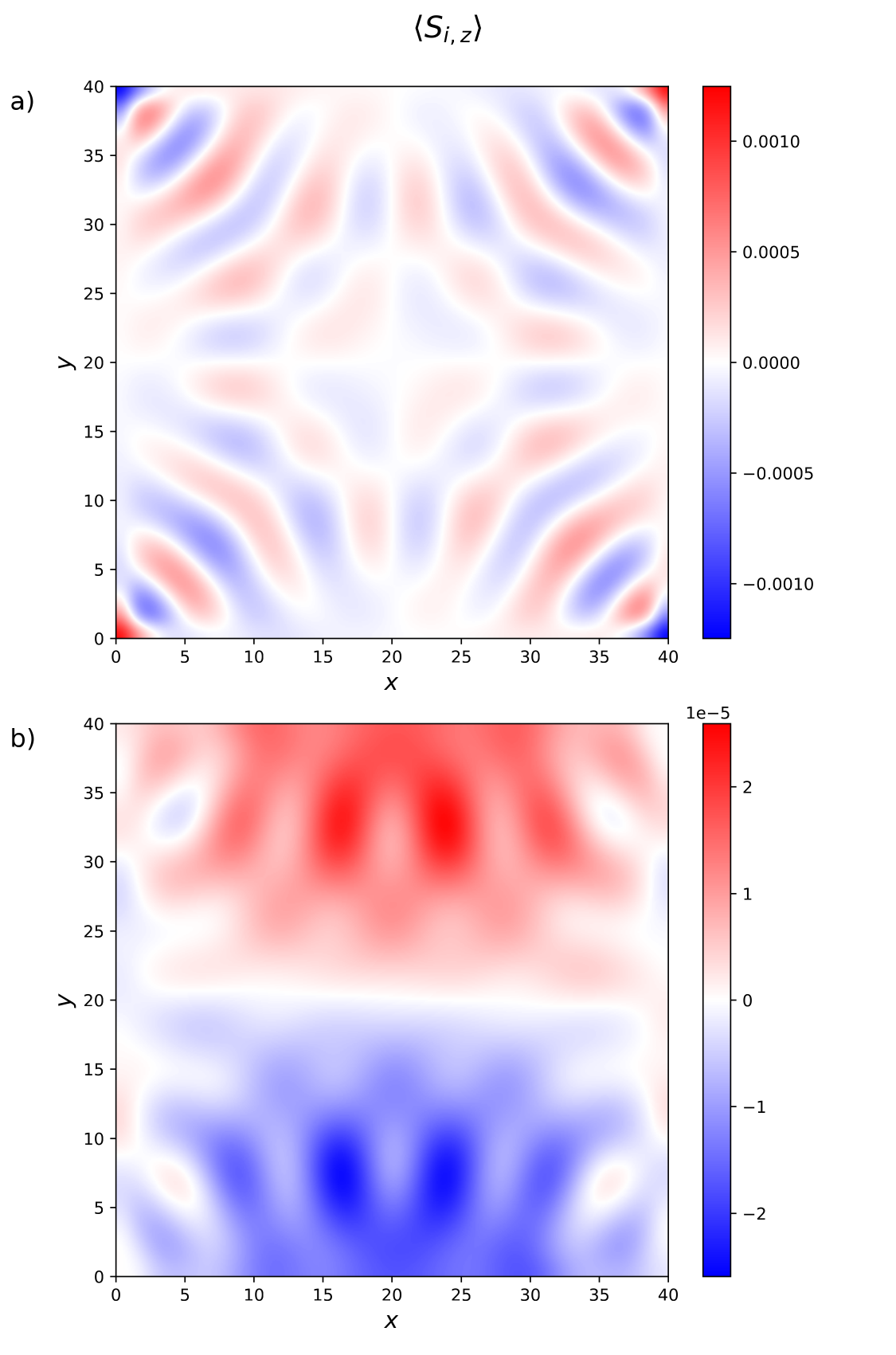}
    \caption{a) Local spin density $\langle \hat S_{\vm}^z \rangle$ for a $40 \times 40$ site $d_{xy}$-altermagnet system with $t_m = 0.1t_S$, $E_F = -3.8t_S$ and applied bias $eV = 0.001t_S$. In the low Fermi-level and bias regime, interface-induced magnetization suppresses the spin accumulation.
    b) Local spin density $\langle \hat S_{\vm}^z \rangle$ for a $40 \times 40$ site $d_{xy}$-altermagnet system with $t_m = 0.1t_S$, $E_F = -3.8t_S$ and applied bias $eV = 0.001t_S$ with the background magnetization removed.}
    \label{fig:40lat0.1am-3.8fermi0.001bias-spindensity}
\end{figure}

Near the edges of the sample, there are clear oscillations in the spin density, especially noticeable on the left and right interfaces. When the applied bias is very small, or the chemical potential very close to the band-bottom, this effect is amplified and eventually leads to a suppression of the spin accumulation. In Fig. \ref{fig:40lat0.1am-3.8fermi0.001bias-spindensity} a) we see that the spin accumulation disappears for a system with $E_F = -3.8t_S$ and bias $eV = 0.001t_S  \ll |E_b - E_F| = 0.2t_S$. We believe this is caused by the spin anisotropy of the altermagnet in combination with Friedel oscillations, as detailed in \cite{hodt2024interface}. In effect, the spin-dependent Fermi vector of the altermagnetic state give rise to spin-dependent Friedel oscillations, whose contribution to the magnetization overshadows the spin-splitter effect in this parameter regime. Further evidence of this may be seen by calculating the local spin density for the same system with no applied bias, and subtracting the result from the biased system. With the background magnetization removed, a spin accumulation pattern arising from the spin-splitter effect again clearly emerges, albeit with a small magnitude. This is shown for the aforementioned system in Fig. \ref{fig:40lat0.1am-3.8fermi0.001bias-spindensity} b). 1D Friedel oscillations, as for instance those associated with the presence of an interface in 2D, generally decay with a length scale set a few Fermi wavelengths $\lambda_\text{F}$,  typically a few nm. As such, the suppression of the spin accumulation should generally be localized to the interfaces, but can be significant in small systems.

\subsubsection{Role of chemical potential}

As we saw in Sec. \ref{subsec:interfacemag}, at chemical potentials close to the band-bottom, the interface-induced magnetization is stronger than the spin-splitter effect. A possible explanation for this is a combination of two effects: At low filling levels, fewer eigenfunctions are occupied compared to higher filling levels. As such, one expects the magnitude of the Friedel oscillations relative to the average filling level to be larger. The magnitude of the interface-induced magnetization generally goes as the difference between the spin-up and -down Friedel oscillations and as such should be proportional to the deviation of these oscillations from an average filling level. These deviations are expected to be relatively larger for lower filling fractions. Secondly, as the chemical potential is reduced and the filling fraction becomes smaller, the DOS is also reduced. With fewer electrons available to provide spin accumulation, the magnitude of the spin-splitter effect diminishes. 

\begin{figure}
    \centering
    \includegraphics[width=0.9\linewidth]{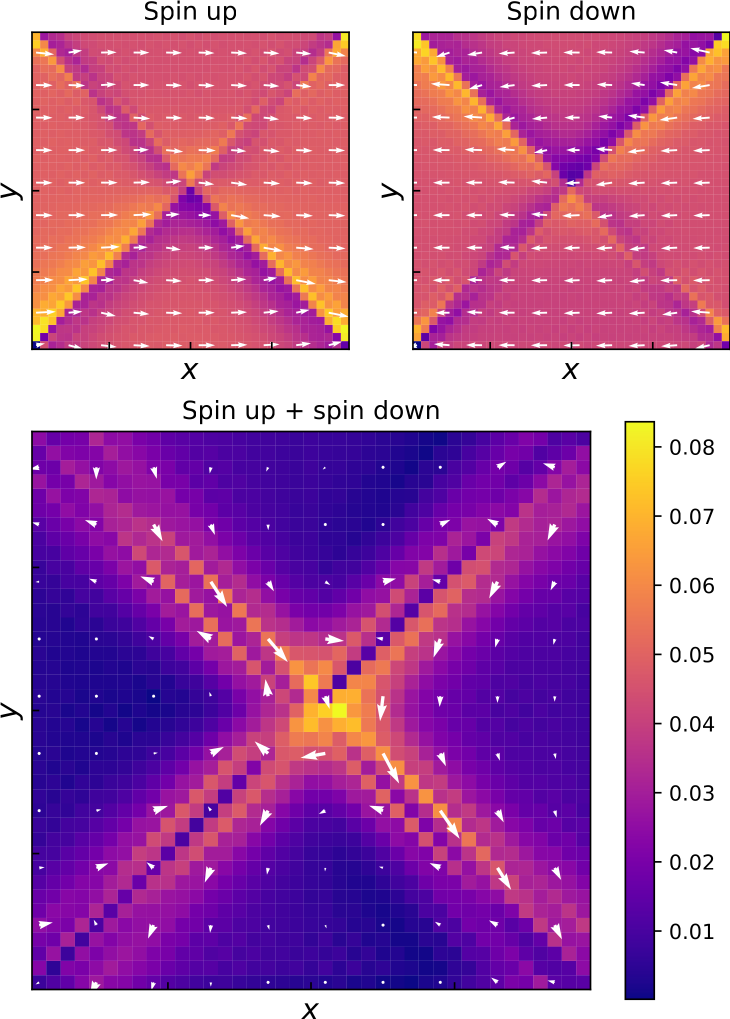}
    \caption{Local spin currents $\langle J_{\vm \vm'}^{\hat S_z}\rangle$ for a $40\times 40$ site $d_{xy}$-altermagnet system with, $t_m = 0.1t_S$, $E_F = -0.0t_S$ and applied bias $eV = 0.4 t_S$. At this filling level, the spin-splitter effect goes to zero as a result of the spin-current changing sign. Additionally, finite size effects due to tight-binding models on square lattices in general add to the cross-structure.}
    \label{fig:40lat0.1am-0.0fermi0.4bias-current}
\end{figure}

New effects emerge for chemical potentials near zero, as seen in Fig. \ref{fig:40lat0.1am-0.0fermi0.4bias-current} for a $40\times40$ system with $E_F = 0$. In addition to the spin-splitter effect being suppressed, there is a noticeable cross-structure appearing in both the magnitude of the local spin currents and the local spin density. 
The latter can be explained by a property of tight-binding models on square lattices, where Friedel oscillations in combination with the shape of the Fermi-surface at half-filling leads to precisely this type of density modulation. The former (vanishing spin-splitter effect) can in fact be proven analytically, as we now proceed to show. In particular, we find that upon reversing the sign of the chemical potential, $\mu \to -\mu$, the spin-splitter effect is identical in magnitude both for current and accumulation, but it changes sign. In other words, our model is particle-hole symmetric if one additionally exchanges spin up and spin down particles. This invariance can be proven as follows.

Consider first a Hamiltonian with only nearest neighbor hopping and a chemical potential, in effect without the altermagnetism. This is the standard non-interacting Hubbard model. This model is known to have particle-hole symmetry under $\mu \to -\mu$. This is confirmed by introducing the transformation $d_{i\sigma}^\dag = (-1)^i c_{i\sigma}$ where $i$ is the site index of our square (bipartite) lattice. The kinetic term, involving nearest-neighbor hopping, is invariant under this transformation since $d_{i\sigma}^\dag d_{j\sigma} = c_{j\sigma}^\dag c_{i\sigma}$ and since there is a summation over all nearest-neighbors $i,j$. The chemical potential term transforms as $d_{i\sigma}^\dag d_{i\sigma} \to 1 - c_{i\sigma}^\dag c_{i\sigma}$. Therefore, this term is only invariant (apart from an irrelevant operator-independent constant) if we let $\mu \to -\mu$. 

We now add altermagnetism. This is modelled in our system through a spin-dependent next-nearest neighbor hopping. When $i,j$ are next-nearest neighbors rather than nearest-neighbors, this affects the transformation of the altermagnetic term. Namely, we obtain 
\begin{align}
    d_{i\sigma}^\dag d_{j\sigma} = -c_{j\sigma}^\dag c_{i\sigma}.
\end{align}
But the altermagnetic next-nearest neighbor hopping has opposite sign for the two spin species. Therefore, the extra sign appearing in the particle-hole transformation can be compensated by simultaneously performing $\sigma \to -\sigma$, in effect swapping spin up and spin down. Therefore, our model, and therefore the spin-splitter effect, is invariant under the combined operation of $\mu\to-\mu$ and $\sigma \to -\sigma$. We have also verified this numerically.

The above observation has interesting consequences with respect to materials where altermagnetism can be described by an effective spin-dependent hopping. Namely, one would expect different behavior for the filling fraction-dependence, or electron- vs hole-doping, depending on whether the altermagnetism arises from nearest- or next-nearest neighbor hopping in a bipartite lattice. If the altermagnetism stems from next-nearest neighbor hopping, it should be particle-hole symmetric, whereas otherwise the electron- and hole-doped regimes should display a different spin-splitter effect. However, particle-hole symmetry may be difficult to realize in realistic antiferromagnetic materials.

\subsubsection{Role of altermagnet strength and Rashba spin-orbit interactions}

\begin{figure}
    \centering
    \includegraphics[width=0.9\linewidth]{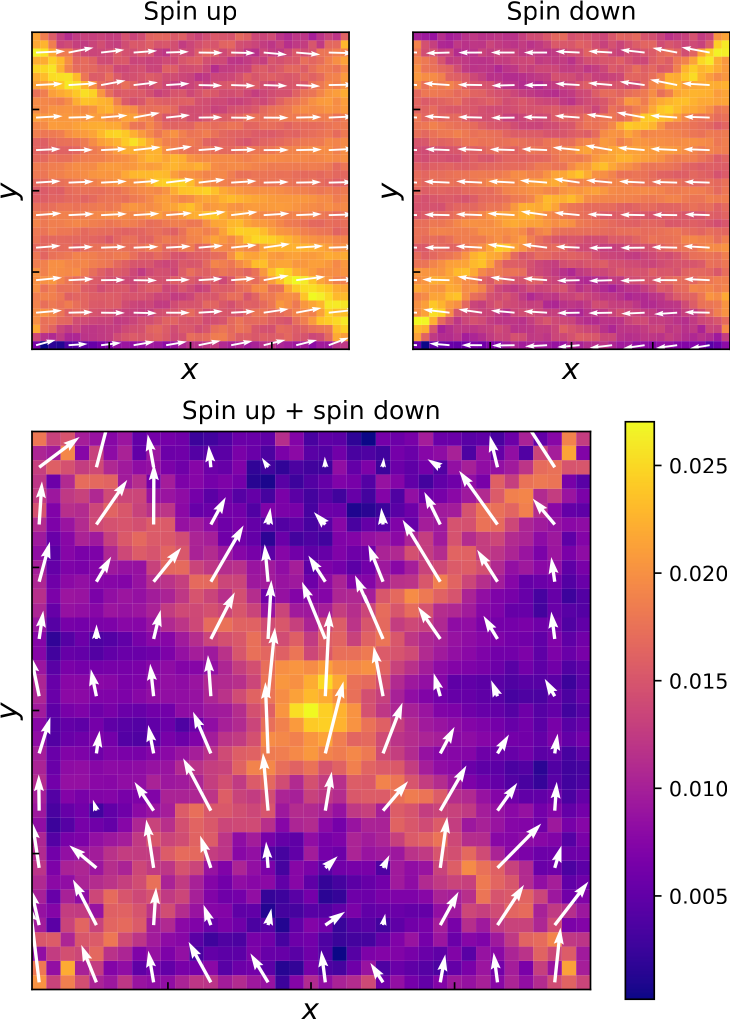}
    \caption{Local spin currents $\langle J_{\vm \vm'}^{S_z}\rangle$ for a $40\times 40$ site $d_{xy}$-altermagnet system of strength $t_m = 0.3t_S$ with reservoirs at Fermi-level $E_F = -2.0t_S$ and applied bias $eV = 0.2t_S$.}
    \label{fig:40lat0.3am-2.0fermi0.2bias-current}
\end{figure}

\begin{figure}
    \centering
    \includegraphics[width=1.0\linewidth]{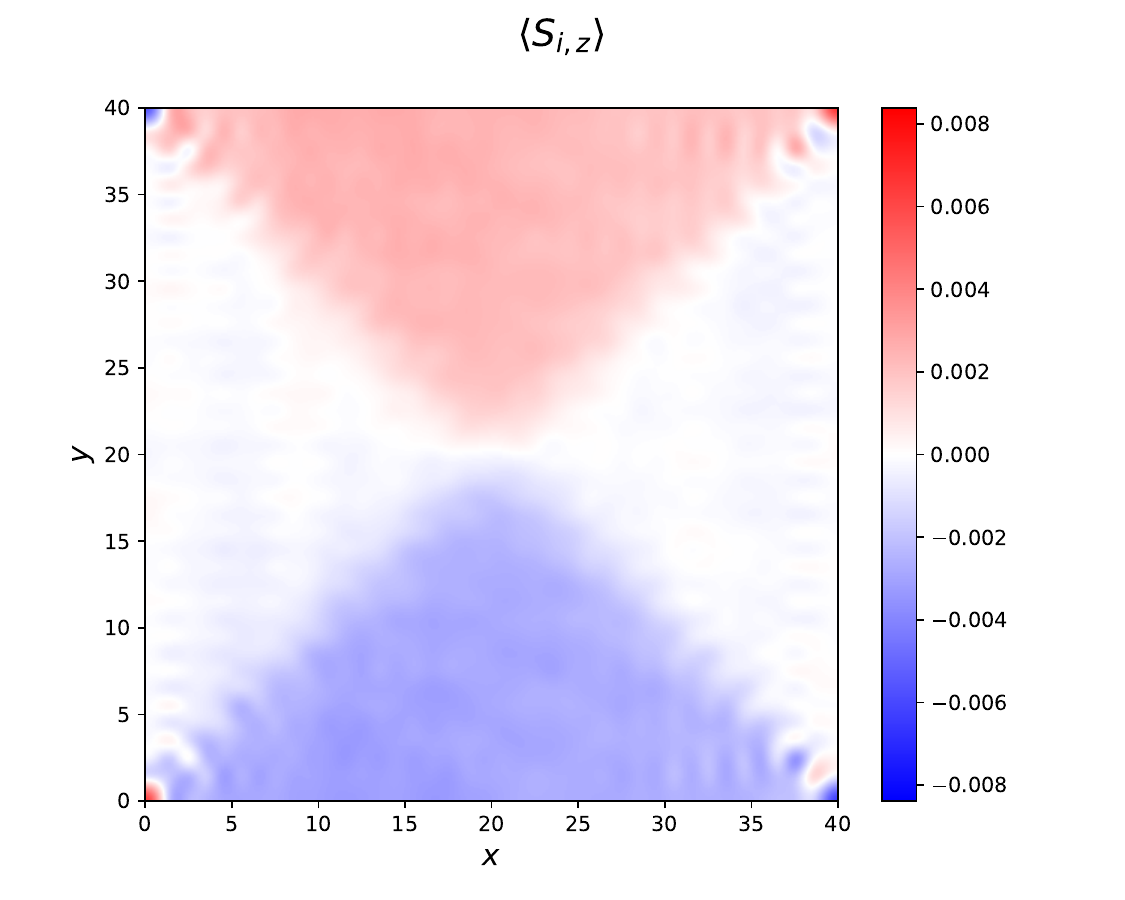}
    \caption{Local spin density $\langle \hat S_{\vm}^z \rangle$ for a $40 \times 40$ site $d_{xy}$-altermagnet system of strength $t_m = 0.3t_S$ with $E_F = -2.0t_S$ and applied bias $eV = 0.2t_S$. At higher altermagnet strength, the accumulation is more pronounced.}
    \label{fig:40lat0.3am-2.0fermi0.2bias-spindensity}
\end{figure}

We have also investigated the effect of modifying the altermagnetic strength in the system. When the strength of the altermagnet-hopping is increased to $t_m = 0.3t_S$, the spin-splitter effect is simply enhanced in magnitude, as expected. This is shown in Figs. \ref{fig:40lat0.3am-2.0fermi0.2bias-current} and \ref{fig:40lat0.3am-2.0fermi0.2bias-spindensity}.

\begin{figure}
    \centering
    \includegraphics[width=1.0\linewidth]{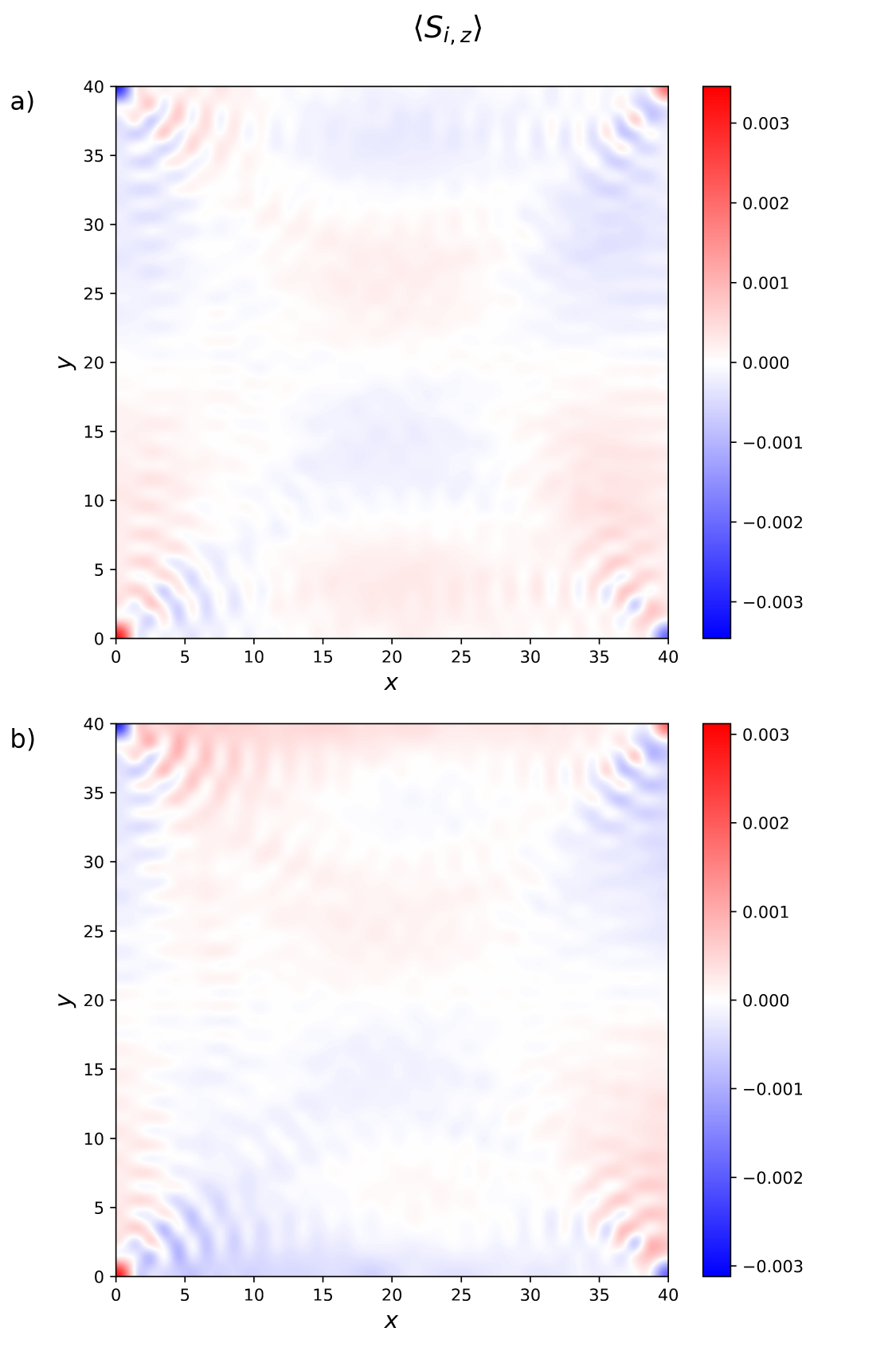}
    \caption{a) Local spin density $\langle \hat S_{\vm}^z \rangle$ for a $40 \times 40$-site $d_{xy}$-altermagnet system with $t_m = 0.1t_S$, $t_{SO} = 0.1t_S$, $E_F = -2.0t_S$ and $eV = 0.2t_S$. For these parameters, spin-orbit coupling is working against the spin-splitter effect.
    b) Local spin density $\langle \hat S_{\vm}^z \rangle$ for a $40 \times 40$-site $d_{xy}$-altermagnet system with $t_m = -0.1t_S$, $t_{SO} = 0.1t_S$, $E_F = -2.0t_S$ and $eV = 0.2t_S$. The sign of the spin accumulation has been flipped to ease comparisons. At these parameter values, the SO-coupling is not working against the spin-splitter effect, but is not linearly increasing it either.}
    \label{fig:40lat0.1am-2.0fermi0.2bias0.1SO-spindensity}
\end{figure}

\begin{figure}
    \centering
    \includegraphics[width=0.9\linewidth]{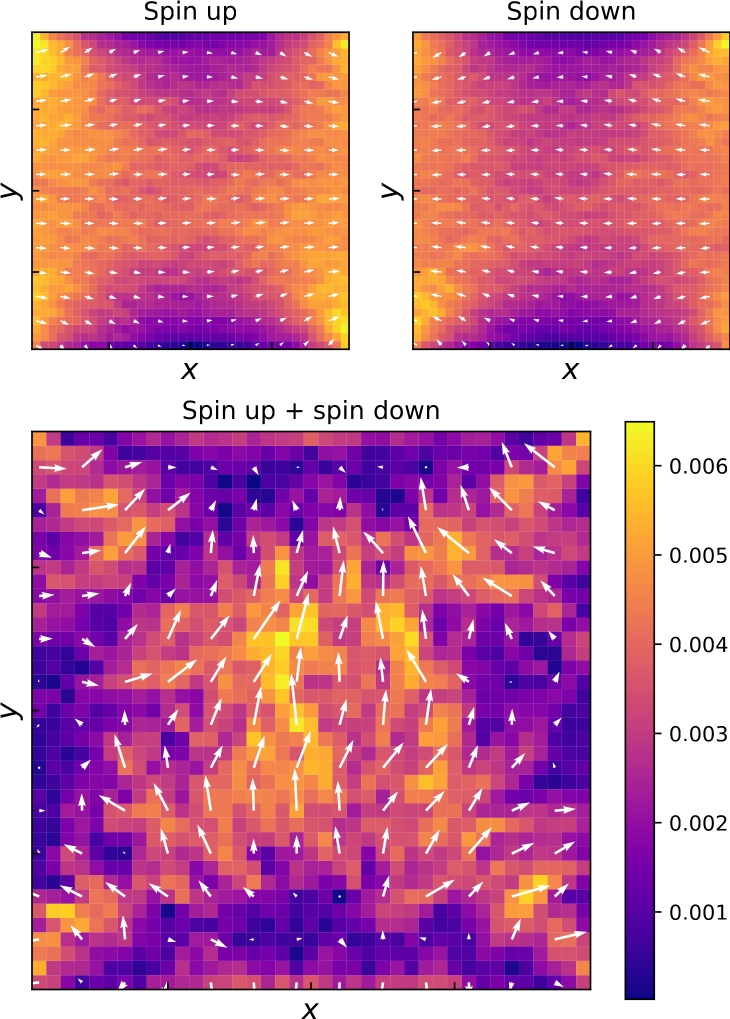}
    \caption{Local spin currents $\langle J_{\vm \vm'}^{\hat S_z}\rangle$ for a $40\times 40$ site $d_{xy}$-altermagnet system with $t_m = 0.1t_S$, $E_F = -2.0t_S$, applied bias $eV = 0.2 t_S$, impurity concentration $n_I = 0.4$ and strength $w = t_S$.}
    \label{fig:40lat0.1am-2.0fermi0.2biasRANDOM401t-current}
\end{figure}
\begin{figure}
    \centering
    \includegraphics[width=1.0\linewidth]{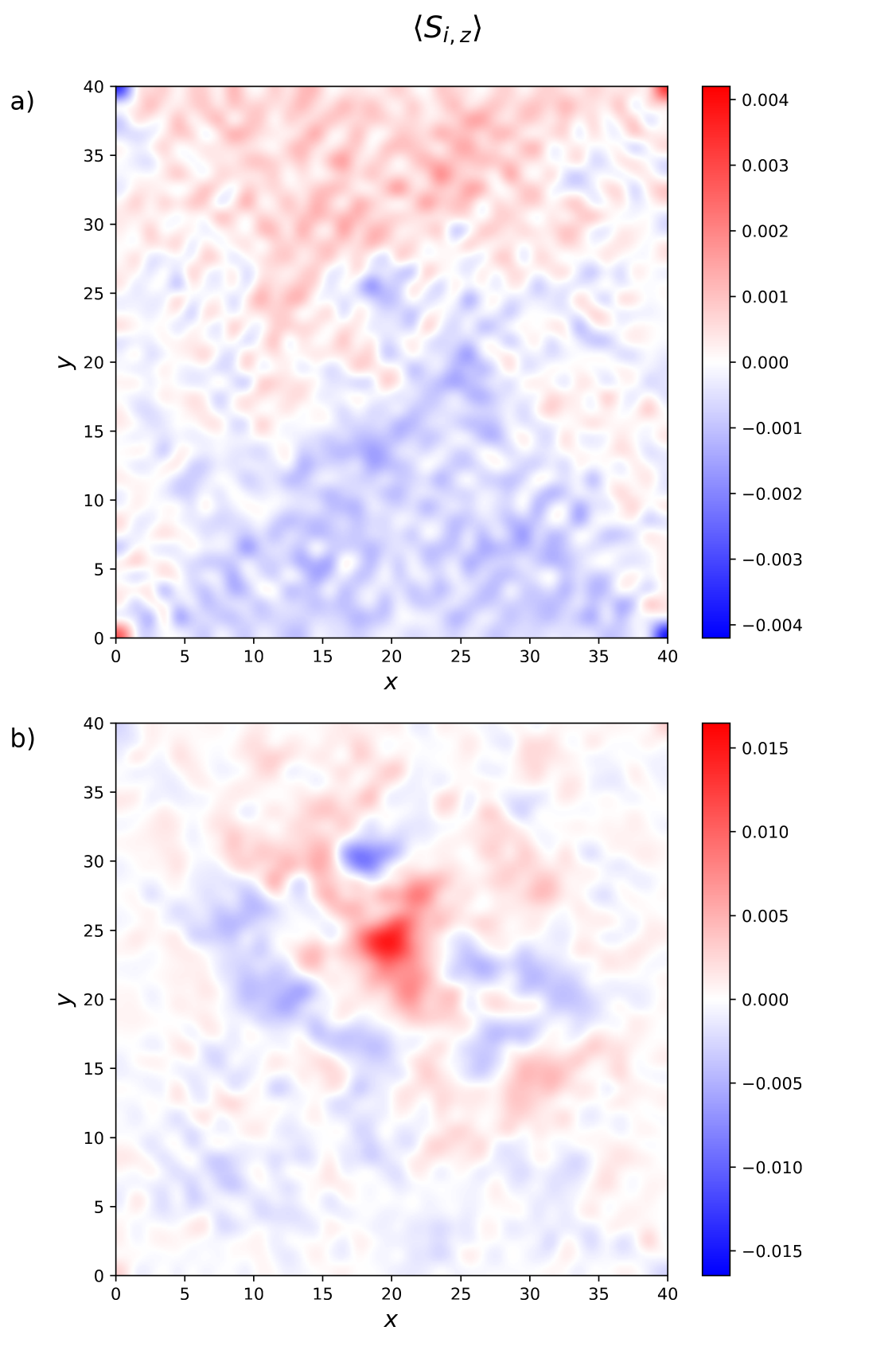}
    \caption{a) Local spin density $\langle \hat S_{\vm}^z \rangle$ for a $40\times 40$ site $d_{xy}$-altermagnet system with $t_m = 0.1t_S$, $E_F = -2.0t_S$, applied bias $eV = 0.2 t_S$, impurity concentration $n_I = 0.4$ and strength $w = t_S$.
    b) Local spin density $\langle \hat S_{\vm}^z \rangle$ for a $40\times 40$ site $d_{xy}$-altermagnet system with $t_m = 0.1t_S$, $E_F = -2.0t_S$, applied bias $eV = 0.2 t_S$, impurity concentration $n_I = 0.4$ and strength $w = 3t_S$.}
    \label{fig:40lat0.1am-2.0fermi0.2biasRANDOM401t-spindensity}
\end{figure}

When spin-orbit interactions are turned on, the spin Hall effect could at first glance be thought to either aid or suppress the spin accumulation caused by the altermagnet linearly. This is because the spin Hall effect also provides a spin accumulation at the transverse edges with a sign depending on the sign of the Rashba spin-orbit coefficient, so that there are two contributions to the spin accumulation: one from the spin-orbit interactions and one from the altermagnetism. However, Fig. \ref{fig:40lat0.1am-2.0fermi0.2bias0.1SO-spindensity} shows that while it is true that spin accumulation is suppressed when the spin-orbit coupling is working against the spin-splitter effect, the spin accumulation is not linearly increased when the signs are working in tandem. We believe that this can be understood at a qualitative level by considering a simplified model for the band structure in the presence of altermagnetism and spin-orbit coupling. Consider the continuum approximation of a $d_{xy}$-altermagnet with spin-orbit coupling \cite{hodt2024interface, am_emerging} given by
\begin{equation}
    \hat H(\v k) = \frac{k^2}{2m} + \alpha \hat \sigma_z k_x k_y + \lambda (\hat \sigma_x k_y - \hat \sigma_y k_x),
\end{equation}
where $k^2 \equiv \v k \cdot \v k$. Here $\alpha$ corresponds to the strength of the altermagnet and $\lambda$ to the spin-orbit coupling. For an altermagnet without spin-orbit coupling, the dispersion relation is given by
\begin{equation}
    E(\vk) = \frac{k^2}{2m} \pm \alpha k_x k_y,
\end{equation}
whereas for a system with only spin-orbit coupling it is given by
\begin{equation}
    E(\vk) = \frac{k^2}{2m} \pm \lambda k,
\end{equation}
with $k = \sqrt{k_x^2 + k_y^2}$. However in a system with both spin-orbit coupling and altermagnets, the dispersion relation is not simply the sum of these, instead it is given by
\begin{equation}
    E(\vk) = \frac{k^2}{2m} + \sqrt{\lambda^2k^2 + \alpha^2k_x^2k_y^2}.
\end{equation}
In effect, since the band structure in the presence of both spin-orbit coupling and altermagnetism is not simply the sum of the corresponding terms of each individually, one should not expect the spin Hall effect and spin-splitter effect to simply add or subtract from each other based on their relative signs. This is in agreement with our numerical simulations.

\subsubsection{Role of impurity scattering}
A recent work derived diffusion equations for spin-flow in dirty altermagnets which describe the spin-splitter effect \cite{kokkeler_arxiv_24} using quasiclassical theory. Here, we provide a fully quantum mechanical description of the spin-splitter effect and describe how it changes when gradually crossing over from the ballistic to an impurity-dominated regime of transport.

To simulate impurities in the central altermagnet region, we uniformly assign a fraction $n_I$ of the sites with a local potential of strength $w$, simulating point-like impurities in a similar manner as Ref. \cite{giil_prb_24}. In Fig. \ref{fig:40lat0.1am-2.0fermi0.2biasRANDOM401t-current} and \ref{fig:40lat0.1am-2.0fermi0.2biasRANDOM401t-spindensity} a), the local spin currents and density are shown for $n_I = 0.4$ and $w = t_S$. Even when a high fraction of the sites contain impurities with a strength comparable to the nearest neighbor hopping strength, the spin-splitter effect is resilient. The local spin current and spin accumulation is still present although weakened, and the presence of the impurities suppress the interface-induced magnetization. At impurity fraction $n_I = 0.4$ and strength $w = 3t_S$, the spin-splitter effect is, however, almost completely suppressed. This is shown in Fig. \ref{fig:40lat0.1am-2.0fermi0.2biasRANDOM401t-spindensity} b) for the local spin density. Our results thus indicate that clean samples are most beneficial with regard to maximizing the spin-splitter signal. These conclusions do not change when considering a lower impurity fraction with higher impurity strength according to our numerical simulations.

\section{Concluding remarks}\label{sec:conclusion}

In this paper we have employed the Keldysh Green function formalism to investigate the spin-splitter effect in altermagnetic materials, where a transverse spin current can be generated even in the absence of spin-orbit coupling. The real-space visualization of spin accumulation induced by the interfaces reveals that altermagnet strength, bias voltage strength, filling level and presence of spin-orbit coupling and impurities affect the magnitude and spatial distribution of the spin accumulation.

Some key findings include the following. The interface-induced magnetization \cite{hodt2024interface} plays a large role in determining if we observe a strong spin accumulation for certain values of the system parameters, in particular its size. The spin-splitter effect vanishes at half-filling in our model due to a combined particle-hole and spin-reversal symmetry. Spin-orbit coupling can suppress accumulation depending on the sign of the altermagnet, but does not linearly increase it when the spin Hall current and spin-splitter current have the same sign. Finally, the presence of moderate impurity scattering with potential strength on the order of the hopping parameter does not completely suppress the spin accumulation, indicating that the effect is robust against disorder. Our work has thus identified under which conditions the spin-splitter effect is most favored, which could be useful as a guide for experiments.

Future research could focus further onto the interplay between the spin-splitter effect and different types of spin-orbit coupling, including spin-orbit impurity scattering, or the phenomena occurring at half-filling for other models of altermagnets or orientations of the electronic band-structure in momentum space. Other types of lattice structures facilitating the same symmetries of the band structure (\ie $d_{xy}$ or $d_{x^2 - y^2}$) should also give rise to a spin-splitter effect, however investigating exactly how the parameters of the model influence the observables in these systems is outside the scope of this work.

\acknowledgments
We thank Pavlo Sukhachov and Niels Henrik Aase for useful discussions. This work was supported by the Research Council of Norway through Grant No. 323766 and its Centres of Excellence funding scheme Grant No. 262633 “QuSpin”. Support from Sigma2 - the National Infrastructure for High-Performance Computing and Data Storage in Norway, project NN9577K, is acknowledged.

\appendix

\section{Reservoir self-energies in the lesser Green function}\label{app:self-energies}
The self-energy terms may be calculated exactly by the procedure detailed in the lecture notes \cite{nikolicselfenergy}, also detailed here for completeness. The leads are modelled using a tight-binding Hamiltonian on a $N^{\mathrm{lead}}_x \times N^{\mathrm{lead}}_y$ lattice where we let \eg $N^{\mathrm{lead}}_x \rightarrow \infty$ for leads situated on the right or left of the sample region. In the case the leads in question are \eg the left or right leads, $N_y^{\mathrm{lead}} = N_y$. For brevity we will omit the superscript indicating these dimensions are those of the leads and not the central region for this section. 

The eigenstates of the lattice are separable into transverse and longitudinal ($y$-direction and $x$-direction respectively for the left and right leads) components in the tight-binding formalism with nearest neighbor hopping and we write the Bloch states as a sum of the orbitals at each site. The Hamiltonian for the bare leads are then given by
\begin{equation}
    \hat H_{\mathrm{lead}} = -t_L \sum_{\langle i, j \rangle} c^\dag_{i}c_j
\end{equation}
With a finite lattice size and open boundary conditions on each side, the eigenstates are given by finding the eigenvectors of the matrix given by
\begin{equation}
    h = \begin{pmatrix}
    0 & -1 & & & 0 \\
    -1 & 0 & -1 & & \\
    & \ddots & \ddots & \ddots & \\
    & & -1 & 0 & -1 \\
    0 & & & -1 & 0
    \end{pmatrix},
\end{equation}
which for brevity is not spin-resolved.
The components $c_n$ of the eigenvector is then found by solving the difference equation
\begin{equation}
    c_{n-1}+Ec_n+c_{n+1}=0, \quad n=1,2,\cdots,N,
\end{equation}
where $c_0 = c_{N + 1} = 0$ by virtue of the states vanishing outside the lattice.
The difference equation may be solved by writing it as
\begin{equation}
    c_{n-1} + c_{n+1} = -Ec_n,
\end{equation}
and guessing the solution is of the form $c_n = \lambda^n$ giving
\begin{equation}
    \lambda^{n-1} + \lambda^{n+1} = -E\lambda^n.
\end{equation}
Dividing by $\lambda^{n-1}$ yields
\begin{equation}
    \lambda^2 + E \lambda + 1 = 0,
\end{equation}
after which we can choose $\lambda = e^{ik}$. This is solved by $E = -2\cos k = -e^{ik} - e^{-ik}$ since
\begin{equation}
    e^{2ik} - (e^{ik} + e^{-ik})e^{ik} + 1 = 0.
\end{equation}
The general solution for $c_n$ is given by
\begin{equation}
    c_n = Ae^{ikn} + Be^{-ikn},
\end{equation}
and from the boundary conditions we can find
\begin{equation}
    c_0 = A + B = 0 \implies A = -B,
\end{equation}
and
\begin{equation}
    c_{N+1} = Ae^{ik(N+1)} - Ae^{-ik(N+1)} = 2iA\sin(k(N+1)) = 0,
\end{equation}
giving conditions on $k$.

For a lattice with lattice constant $a$, the components are found to be $c_n = C\sin (k n a)$ where $k$ is quantized via
\begin{equation}
    k(j) = \frac{\pi}{a(N + 1)}j, \quad j = 1, \dots, N,
\end{equation}
and $C$ is a normalization constant. This gives an expression for \eg the normalized left and right lead transverse eigenstates as
\begin{equation}
    |k_y \rangle = \sqrt{\frac{2}{N_y + 1}}\sum_{n_y = 1}^{N_y}\sin{(k_y n_y a)}|n_y \rangle.
\end{equation}
For normal metal reservoirs, the eigenvalues are given by
\begin{equation}
    \epsilon(k_y) = 2t_L \cos (k_y a).
\end{equation}
To model reservoirs with different properties, one would modify this dispersion relation and distribution function of the reservoirs that appear in Eq. (\ref{sigmalesser}).
Here $|n_y \rangle$ denotes the orbital of the lattice site with $y$-index $n_y$, running from $1$ to $N_y$.

The longitudinal eigenstates of the bare leads (which vanish on the edge $n_x = 0$ and at infinity) has components given by
\begin{equation}
    \langle n_x | k_x \rangle = \sqrt{\frac{2}{N_x^{\mathrm{lead}}}}\sin(k_x n_x a), 
\end{equation}
with non-quantized eigenvalues 
\begin{equation}
    \epsilon(k_x) = 2t_L \cos(k_x a).
\end{equation}
The Green function on the sites corresponding to the edge of the sample and leads may then be expanded in terms of these eigenstates via
\begin{equation}\label{baregreens}
\begin{split}
    &\langle m_x, m_y | g^R_{\mathrm{lead}} | m'_x, m'_y \rangle \\
    &= \sum_{k_x, k_y} \langle m_y | k_y \rangle \langle k_y | m'_y \rangle \frac{2}{N_x^{\mathrm{lead}}} \frac{\sin^2 (k_x a)}{E - \epsilon(k_y) - 2t_L \cos(k_x a) + i\eta},
\end{split}
\end{equation}
where $\epsilon(k_y) + 2t_L \cos(k_xa)$ are the energy eigenvalues of the lead Hamiltonian $\hat H_L$ and the bare Green function is given by $g^R_{\mathrm{lead}} = (E - \hat H_L + i\eta)^{-1}$. Since we assume hopping only between the leads and the closest layer on the lead only the term with $m'_x = 1$ for \eg the left lead contributes.

When letting $N_x^{\mathrm{lead}} \rightarrow \infty$, $k_x$ becomes continuous and the sum over $k_x$ may be exchanged for an integral via
\begin{equation}
\begin{split}
    J(k_y) \equiv \frac{2}{N^{\mathrm{lead}}_x}\sum_{k_x}\frac{\sin^2(k_x a)}{E - \epsilon(k_y) - 2t_L \cos(k_xa) + i\eta} \\
    = \frac{a}{4\pi t_L}\int_0^{\pi/a}dk_x \frac{2 - e^{2ik_xa} - e^{-2ik_xa}}{(E_J + i\eta)/2t_L - \cos(k_x a)},
\end{split}
\end{equation}
where $E_J \equiv E - \epsilon(k_y)$. The integral may be solved by rewriting the integral into a complex integral over the unit circle and using the residue theorem.
\begin{equation}
    J(k_y) = -\frac{1}{4i\pi t_L}\oint_{|w|=1}\frac{1 - w^2}{w^2/2 + 1/2 - Yw}, 
\end{equation}
with $Y \equiv (E_J + i \eta)/2t_L$. If $|Y| \leq 1$, both of the poles are on the unit circle and the $+i\eta$ is needed to uniquely define the retarded Green function. In this case, the integral evaluates to
\begin{equation}
    J(k_y) = \frac{1}{2t_L^2}\big(E_J - i\sqrt{E_J^2 - 4t_L^2}\big),
\end{equation}
whilst if $|Y| > 1$ it evaluates to
\begin{equation}
    J(k_y) = \frac{1}{2t_L^2}\big(E_J - \mathrm{sign}(E_J)\sqrt{E_J^2 - 4t_L^2}\big).
\end{equation}
Inserting these expressions into Eq. (\ref{baregreens}) then yields the two cases for \eg $\vm, \vm'$ corresponding to the left edge of the sample
\begin{equation}
\begin{split}
    \Sigma_{L,\vm \vm'} = \frac{2}{N_y + 1}\sum_{k_y}\sin(k_y m_y a)\sin(k_y m'_y a) \\
    \times \frac{t_C^2}{2t_L^2}\big(E_J - i\sqrt{4t_L^2 - E_J^2}\big),
\end{split}
\end{equation}
when $|E_J| \leq 2t_L$ and
\begin{equation}
\begin{split}
    \Sigma_{L, \vm \vm'} = \frac{2}{N_y + 1}\sum_{k_y}\sin(k_y m_y a)\sin(k_y m'_y a) \\
    \times \frac{t_C^2}{2t_L^2}\big(E_J - \mathrm{sign}(E_J)\sqrt{E_J^2 - 4t_L^2}\big),
\end{split}
\end{equation}
for $|E_J| > 2t_L$. For the leads situated above and below the sample region the expressions are similar with the exception that the relevant lattice sites are the uppermost or lowermost horizontal row of the sample region, and hence instead of using $m_y$, $m_y'$ and $N_y$ one should use $m_x$, $m_x'$ and $N_x$.

\section{Derivation of Keldysh equation}\label{sec:keldysh}
We follow here partially the derivation in Ref. \cite{maciejko2007nonequilibrium}. The starting point is the contour Dyson equation on the Schwinger-Keldysh contour:
\begin{align}
    G(1,1') = G_0(1,1') + \int d2 \int d3 G_0(1,2) \Sigma(2,3) G(3,1')
\end{align}
where $G(1,1') = -i T_c\langle \psi(1)\psi^\dag(1')\rangle$ is the exact contour-ordered Green function, whereas $G_0(1,1')$ is the unperturbed Green function. The field operators are in the interaction picture whereas $\Sigma$ is a 1-particle irreducible self-energy. The integration is short-hand notation for a sum over all internal variables, so that for instance $\in d2 = \sigma_{\sigma_2} \int d\vecr_2 \int_C d\tau_2$. To shorten the notation further, we write the above equation simply as
\begin{align}
    G = G_0 + G_0\Sigma G.
\end{align}
There exists a set of theorems for 'matrix products' consisting of convolution integrals on the contour $C$. Consider the product:
\begin{align}
    C(1,1') = \int d2 A(1,2) B(2,1')
\end{align}
which we would write as $C=AB$ in short-hand notation. The Langreth theorem then states that
\begin{align}
    (AB)^< &= A^RB^< + A^< B^A \notag\\
    (AB)^{R,A} &= A^{R,A}B^{R,A}\\
    (ABC)^< &= A^RB^RC^< + A^RB^<C^A + A^<B^AC^A\notag\\
    (ABC)^{R,A} &= A^{R,A}B^{R,A}C^{R,A}.
\end{align}
According to the first rule applied on the contour Dyson equation, we obtain
\begin{align}\label{eq:firstddysonGR}
    G^R = G_0^R + G_0^R\Sigma^R G^R.
\end{align}
Since the Dyson equation for the contour-ordered Green function can also be written as $G = G_0 + G\Sigma G_0$, we also have that 
\begin{align}\label{eq:seconddysonGR}
    G^R = G_0^R + G^R\Sigma^R G_0^R.
\end{align}
Applying the third Langreth rule above the contour Dyson equation, we obtain
\begin{equation}
    G^< = G_0^< + G_0^R\Sigma^RG^< + G^R_0 \Sigma^< G^A + G_0^< \Sigma^A G^A.
\end{equation}

Now subtract $G_0^R\Sigma^RG^<$ from both sides to get
\begin{equation}\label{eq:Gintermediate}
    (1 - G_0^R\Sigma^R)G^< = G_0^<(1 + \Sigma^A G^A) + G_0^R \Sigma^< G^A.
\end{equation}
Use the Dyson equation for $G^R$ in Eq. (\ref{eq:firstddysonGR})
\begin{equation}
    G^R = G_0^R + G_0^R\Sigma^R G^R \implies (1 - G_0^R\Sigma^R)G^R = G_0^R,
\end{equation}
and multiply on the right by $(G^R)^{-1}$
\begin{equation}
    (1 - G_0^R\Sigma^R) = G_0^R (G^R)^{-1}
\end{equation}
Substitute this into Eq. (\ref{eq:Gintermediate}) to get
\begin{equation}
    G_0^R (G^R)^{-1}G^< = G_0^<(1 + \Sigma^A G^A) + G_0^R \Sigma^< G^A.
\end{equation}
Now multiply from the left by $G^R(G_0^R)^{-1}$ which yields
\begin{equation}
    G^< = G^R (G_0^R)^{-1} G_0^<(1 + \Sigma^A G^A) + G^R \Sigma^< G^A.
\end{equation}
However, for the non-perturbed Green function, the relation
\begin{align}
    (G_0^R)^{-1} G_0^<=0
\end{align}
holds. Therefore, we end up with simply
\begin{align}
G^< = G^R\Sigma^< G^A.
\end{align}

\bibliography{ref}

\end{document}